\documentclass[sigconf]{acmart}
\usepackage{xcolor}
\usepackage{colortbl}
\usepackage{tcolorbox}
\usepackage{listings}
\usepackage{makecell}
\usepackage{multirow}
\usepackage{amsmath,amsfonts}

\usepackage{amssymb}
\usepackage{algorithmic}
\usepackage{graphicx}
\usepackage{fbox}
\usepackage{textcomp}
\usepackage{minibox}
\usepackage{balance}
\usepackage{enumitem}
\usepackage{xspace}
\usepackage[T1]{fontenc}
\usepackage{booktabs}
\usepackage{threeparttable}
\usepackage[]{collab}
\usepackage{url}
\usepackage{hyperref}
\usepackage{natbib}

\usepackage{threeparttable}

\usepackage{subfigure}
\usepackage{arydshln}
\usepackage[ruled,linesnumbered]{algorithm2e}
\usepackage{cleveref}
\usepackage{pifont}

\AtBeginDocument{%
  \providecommand\BibTeX{{%
    \normalfont B\kern-0.5em{\scshape i\kern-0.25em b}\kern-0.8em\TeX}}}

\copyrightyear{2024} 
\acmYear{2024} 
\setcopyright{rightsretained} 
\acmConference[ASE '24]{39th IEEE/ACM International Conference on Automated Software Engineering }{October 27-November 1, 2024}{Sacramento, CA, USA}
\acmBooktitle{39th IEEE/ACM International Conference on Automated Software Engineering (ASE '24), October 27-November 1, 2024, Sacramento, CA, USA}
\acmDOI{10.1145/3691620.3695503}
\acmISBN{979-8-4007-1248-7/24/10}

\newcommand{\mine}{CoCoMine\xspace}
\newcommand{\model}{\textsc{DataCoder}\xspace}
\newcommand{\dataset}{CoCoNote\xspace}
\def\dsp/{\textsc{DSP}}

\newcommand{\ie}{{\em i.e.},\xspace}
\newcommand{\eg}{{\em e.g.},\xspace}

\definecolor{ballblue}{rgb}{0.13, 0.67, 0.8}
\definecolor{jcpink}{RGB}{255, 0, 96}
\collabAuthor{jj}{purple}{Junjie}
\collabAuthor{cl}{red}{Chenglong}
\collabAuthor{jz}{magenta}{Jiazhen Gu}

\begin{document}
\title{Contextualized Data-Wrangling Code Generation in Computational Notebooks}

\author{Junjie Huang$^1$, Daya Guo$^2$, Chenglong Wang$^3$, Jiazhen Gu$^1$,\\ Shuai Lu$^4$, Jeevana Priya Inala$^3$, Cong Yan$^3$, Jianfeng Gao$^3$,  Nan Duan$^4$,  Michael R. Lyu$^1$}
\affiliation{%
  \institution{$^1$The Chinese University of Hong Kong, China.  Email: \{jjhuang23, jiazhengu, lyu\}@cse.cuhk.edu.hk}
  \country{}}
\affiliation{%
  \institution{$^2$Sun-yat Sen University. Email: guody5@mail2.sysu.edu.cn}
  \country{}}
\affiliation{%
  \institution{$^3$Microsoft Research. $^4$Microsoft Research Asia. Email: \{chenglong.wang, shuailu, jinala, jfgao\}@microsoft.com}
  \country{}}

\renewcommand{\shortauthors}{Junjie Huang, et al.}

\begin{abstract}
Data wrangling, the process of preparing raw data for further analysis in computational notebooks, is a crucial yet time-consuming step in data science. Code generation has the potential to automate the data wrangling process to reduce analysts' overhead by translating user intents into executable code.
Precisely generating data wrangling code necessitates a comprehensive consideration of the rich context present in notebooks, including textual context, code context and data context. However, notebooks often interleave multiple non-linear analysis tasks into linear sequence of code blocks, where the contextual dependencies are not clearly reflected. Directly training models with source code blocks fails to fully exploit the contexts for accurate wrangling code generation.

To bridge the gap, we aim to construct a high quality datasets with clear and rich contexts to help training models for data wrangling code generation tasks. In this work, we first propose an automated approach, \mine to mine data-wrangling code generation examples with clear multi-modal contextual dependency. It first adopts data flow analysis to identify the code blocks containing data wrangling codes. Then, \mine extracts the contextualized data-wrangling code examples through tracing and replaying notebooks. 
With \mine, we construct \dataset, a dataset containing 58,221 examples for \emph{\textbf{Co}ntextualized Data-wrangling \textbf{Co}de generation in \textbf{Note}books}. 
To demonstrate the effectiveness of our dataset, we finetune a range of pretrained code models and prompt various large language models on our task. 
Furthermore, we also propose \model, which encodes data context and code\&textual contexts separately to enhance code generation. Experiment results demonstrate the significance of incorporating data context in data-wrangling code generation and the effectiveness of our model. We release code and data at \url{https://github.com/Jun-jie-Huang/CoCoNote}.

\end{abstract}


\maketitle

\section{Introduction}\label{sec:intro}

Data wrangling involves cleaning, structuring, and enriching raw data into a desired format for further analysis~\cite{yan2020autosuggest}, such as by removing duplicates, casting types, and extracting features~\cite{chopra2023cowrangler}. This procedure is notably time-consuming, with data scientists reportedly spending up to 80\% of their time on repetitive data wrangling scripts~\cite{Gulwani2016PBEData, chopra2023cowrangler, dasu2002miningDB}. Consequently, automating the generation of data wrangling codes is crucial for improving the productivity of data science practitioners~\cite{bavishi2019autopandas, yan2020autosuggest}.

\begin{figure}[t]
    \centering
    \includegraphics[width=0.99\columnwidth]{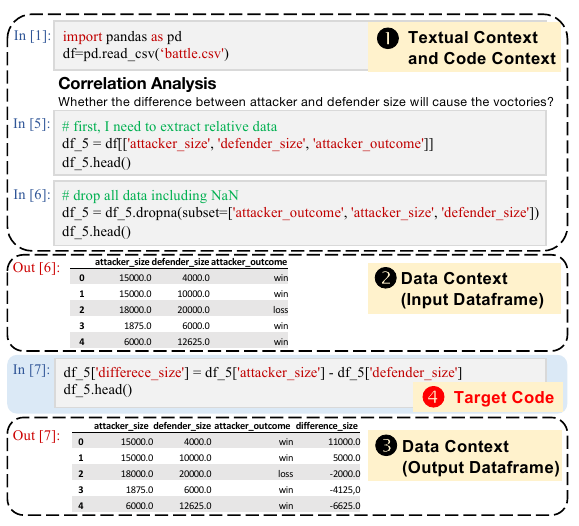}
    \caption{An example of contextualized data-wrangling code generation in \dataset, aiming to generate target code(\ding{185}) given multi-modal context(\ding{182}\ding{183}\ding{184}).}
    \label{fig:intro:task-definition}
    \vspace{-6mm}
\end{figure}

Data wrangling code is typically written in computational notebooks, such as Jupyter Notebooks~\cite{Kluyver2016JupyterNotebook} and Google Colab~\cite{bisong2019googleColab}, rather than in traditional editors or integrated development environments (IDEs). These notebooks allow users to write code and annotations in \textit{cells}, enabling interactive execution of code, inspection of intermediate outputs, and planning of further data analysis~\cite{rule2018exploration, Wang2021AutoDSTH}. As a result, the notebook environment provides rich, multi-modal contextual information, including code, textual elements (\eg code comments and markdown text), and data~\cite{wang2023supernova}, as illustrated in Figure~\ref{fig:intro:task-definition}.

Precise data-wrangling code generation necessitates fully exploiting all contextual information (\ie code, textual and data). Existing solutions often fail to integrate these contexts effectively, leading to suboptimal performance.
Specifically, several studies have focused on generating code based on textual and code context~(\ie \ding{182}$\rightarrow$\ding{185} in Figure~\ref{fig:intro:task-definition})~\cite{agashe2019juice, Yin2022DSN, Chandel2022JuPyT5, Huang2022ExeDS}. However, the inherent ambiguity of natural language~\cite{li2023starcoder} and its potential absence in notebooks~\cite{mondal2023cell2doc} (\eg the textual context fails to provide explicit instructions for the target code in Figure~\ref{fig:intro:task-definition}) can make the specification unclear. 
Another approach adopted the programming-by-example (PBE) paradigm~\cite{Gulwani2016PBEData}, where data context (\ie input and output examples) serves as specifications~(\ie \ding{183}$+$\ding{184}$\rightarrow$\ding{185})~\cite{chen2021latent, barowy2015flashrelate, parisotto2017flashfill}. However, the absence of codes and annotations can lead to a lack of preceding programming details, resulting in erroneous generated codes, such as misused APIs or variables.
Additionally, some studies~\cite{jain2022jigsaw, khatry2023words2code} used textual and code context alongside input examples~(\ie \ding{182}$+$\ding{183}$\rightarrow$\ding{185}). Nevertheless, without the output data, crucial information may still be missing. For instance,  as shown in Figure~\ref{fig:intro:task-definition}, the data field \texttt{difference\_size} never appears in either the textual context or the input example, leaving key details unspecified.
Therefore, integrating code, textual, and data contexts is essential to fully elucidate specifications for data-wrangling code generation tasks.

Ideally, we aim to train a model capable of comprehending all three types of contexts to generate data wrangling code. However, the primary challenge in building such a model is the scarcity of high-quality datasets. Notebooks often interleave multiple non-linear analysis tasks into a linear sequence of code blocks, obscuring contextual dependencies. Consequently, models trained solely on source code fail to fully utilize these contexts for generating wrangling code. This issue is further compounded by the fact that many online notebooks lack complete context information, such as missing execution histories that are crucial for providing essential data contexts.

This work aims to construct a high quality dataset with clear and rich contexts to facilitate training models for data wrangling code generation tasks.
We propose an automated approach, \ie \mine (Contextualized data-wrangling Code Mining), to extract data-wrangling code examples with multi-modal contexts.
\mine consists of two major steps: data-flow-based data-wrangling code identification and multi-modal contexts alignment.
First, \mine extracts relevant code cells for data analysis (including data initialization, wrangling, and utilization~\cite{biswas2022dsp}) from the tangled notebooks.
To identify data-wrangling code cells, we create API databases to signify the data initialization and utilization stages, where the dataframe variable is initially created and finally utilized, such as visualization and statistic computation. \mine then conducts data-flow analysis on the dataframe variable to track its intermediate transformation process, treating all related code cells as data-wrangling cells. 
With the identified code-wrangling cells, \mine creates contextualized code generation examples.
In particular, it first collects the code and textual context through dependency analysis. Then, \mine gathers the intermediate execution results of the dataframe variable as data context by crawling imported data files and replaying the notebook in a sandbox environment. Finally, the target code, along with its aligned code, textual, and data contexts, form a comprehensive code generation example. 

By employing \mine, we collected a dataset comprising 58,221 examples (54,567 for training, 2,000 for development, and 1,654 for testing) for \emph{\textbf{Co}ntextualized \textbf{Co}de generation in \textbf{Note}books}, named \dataset. To ensure the natural and reliable code evaluation, \dataset can be evaluated not only with surface-form metrics such as Exact Match (EM) and CodeBLEU~\cite{ren2020codebleu}, but also with an execution-based metric named Execution Accuracy (EA). The EA evaluates the generated code by comparing the resulting output dataframes against the oracle output, similar to the approaches in~\cite{Yin2022DSN, Huang2022ExeDS}.

With \dataset, we conduct an extensive study on the impact of multi-modal context and the performance of various language models on data-wrangling code generation. 
Specifically, we used \dataset to fine-tune five code-pretrained language models (PLMs) and prompt six large language models (LLMs) to comprehensively evaluate their performance.
During model fine-tuning stage, we observed that data context, which includes structured dataframes, should not be encoded alongside unstructured code and textual contexts due to their differing nature. Additionally, the input length limitation of PLMs make it impractical to encode all three contexts simultaneously. 
To fully leverage data contexts, we extend CodeT5~\cite{wang2021codet5} and proposed a novel encoder-decoder model called \model. \model employs dual encoders to separately encode the code and textual contexts and input-output dataframes, resulting in more accurate latent representations for decoding. 

Our experimental results demonstrate that (1) multi-modal contexts, particularly the data context of input-output examples, are crucial for data-wrangling code generation, as each modality provides essential information for model inference. 
(2) Column names in data context have a greater impact than data values in offering explicit signals for data manipulation, while output examples are more effective than input examples in clarifying the intended results by specifying the expected output. 
(3) With increased parameters and enhanced training methods, GPT-4 achieves the highest performance with an execution accuracy of 50.6\%, indicating substantial room for further improvement on this task. We hope this study will benefit future research in developing robust multi-modal code generation methods.

We summarize the contributions as follows: 
\begin{itemize}[leftmargin=*]
    \item We introduce the task of contextualized data-wrangling code generation, incorporating code context, textual context and data context.
    \item We develop \mine, a tool to mine comprehensive context for data-warngling codes, resulting in a dataset of 58,221 examples.
    \item We propose a data-enhanced model, \model, featuring separate encoders for data and textual/code, which better utilizes contextual information to improve code generation.
    \item We provide a comprehensive analysis of how contextual information affects generation performance across a range of PLMs and LLMs. 
\end{itemize}

\section{Task and Background}\label{sec:task-define}

This paper aims to explore the role of various context in data-wrangling code generation under the notebook environment. We narrow down our focus to data wrangling, the most time-consuming and tedious process in data analysis~\cite{Gulwani2016PBEData, Wang2021AutoDSTH, yan2020autosuggest}, where the data scientist's target is to manipulate data to the right shape in preparation for for subsequent analysis including plotting, modeling or statistics computation~\cite{bavishi2019autopandas}. 

\subsection{Motivation}\label{sec:task-define-motivation}
Our work is motivated by a common programming scenario in which data scientists program in computational notebooks to analyze the target data. As they program, DS practitioners often consult various sources of information within notebooks to help them write correct analysis code~\cite{wang2023supernova}, \eg code, markdown text, tabular dataframe, images, and external file system~\cite{wang2023supernova}. 
Code snippets, written in preceding code cells, provide essential code dependencies such as imported modules, methods, and variable names~\cite{agashe2019juice}. Text in markdown cells or inline comments can provide valuable instructions to the task background and intended output~\cite{nijkamp2022codegen}. 
Apart from code and text, tabular dataframe is another source that data scientists might refer to. By looking up dataframes, they can understand column names and value types, assisting data wrangling process~\cite{chen2021plotcoder}. 
Hence, comprehensive notebook contexts in multi-modalities are important to write correct data-wrangling code.

\subsection{Task Formulation} \label{sec:task-define-formulation}
To simulate the DS programming practice and investigate the effects of multimodal programmatic context, we formulate a novel task of \textit{contextualized data-wrangling code generation} in notebooks. As shown in Figure~\ref{fig:intro:task-definition}, given a mix of code and textual context and data context (\ie input-output dataframes), the task aims to generate target code that can transform the input dataframe to the output dataframe. In this task, the code and context contexts come from the preceding notebook cells that are relevant to target code. The input dataframe is the runtime data to manipulate, which is usually displayed for perceiving, while the output dataframe is a sample of execution result, which specify the expected output. 
We study \textit{dataframe}, a standard data structure to store tabular data in \texttt{Pandas} library, due to its popularity in data analytics and powerful data manipulation capabilities~\cite{pandas}.

Unlike traditional code generation in notebooks~\cite{Huang2022ExeDS, Chandel2022JuPyT5,Yin2022DSN} that includes a textual intent to specify target code function, we provide input-output data as the instruction to constrain the task. This novel setting resembles a programming-by-example (PBE)~\cite{Pattis1994Karel} fashion and is motivated by three factors. 
Firstly, NL isn’t always the optimal specification type, as different data transformation tasks may require different specification types~\cite{wang2023dataformulator}. For example, “pivot” tasks are hard to describe precisely in NL, whereas “sort” tasks are more suitable for NL~\cite{wang2023dataformulator}. 
Secondly, textual intents have inherent ambiguity in expressing these tasks due to the gap between requirements and program logic~\cite{yin2022ingredients}. Whereas concrete input-output examples can elucidate specifications and have been observed and verified as a useful means to provide task intents~\cite{Gulwani2016PBEData}. 
Thirdly, writing complete textual specification is hard and time-consuming for programmers even with extensive expertise~\cite{Gulwani2016PBEData }. In contrast, input-output data supplement data wrangling by specifying expected outcomes and thus enable non-programmers to solve this task. 
Therefore, our task might potentially revolutionize automated data wrangling since users can write the requirements using examples, as discussed in previous study~\cite{gulwani2012spreadsheet, meng2013lase}.

Our work also distinguishes from existing data-wrangling code synthesis approaches that rely solely on input-output examples~\cite{bavishi2019autopandas, yan2020autosuggest} or incorporate additional textual intent~\cite{jain2022jigsaw, khatry2023words2code} by providing more natural and comprehensive programmatic contexts. 
These contexts provide essential references, such as variables to process and code patterns~\cite{chen2021plotcoder}. Our experiment results (\S\ref{sec:experiment-context}) also show the importance of these programmatic context in generating executable and correct code. 
Generally, our task brings PBE to a more practical scenario~\cite{chen2021latent} in computational notebooks by augmenting programmatic context.

\subsection{Evaluation:} \label{sec:task-define-evaluation}
To assess the correctness of generated code, we leverage both surface-form metrics and execution-based metrics~\cite{chen2021evaluating} for rigorous and robust evaluation in our task. For surface-form metrics, we use two prevailing metrics, \ie \textbf{Exact Match (EM)} accuracy and \textbf{CodeBLEU (CB)}~\cite{ren2020codebleu}, to measure the surface correctness. 
The EM accuracy is the most strict metric that measures whether the code generated by the model is identical to the ground-truth code.
CB takes both the code structure and semantics into account to measure semantic and syntactic similarity of two code snippets. It is computed as the average of four components including n-gram matching score, weighted n-gram matching score, syntactic AST matching score, and semantic data flow matching score.
For execution-based metrics, we create a sandbox environment to execute generated code and use \textbf{Execution Accuracy (EA)}, the percentage of examples that generated and ground-truth code produces the same output dataframe, to measure the execution correctness. The sandbox environment setup is elaborated in Section~\ref{sec:dataset-cge-target-code}.

\begin{figure*}[t]
    \centering
    \includegraphics[width=0.9\linewidth]{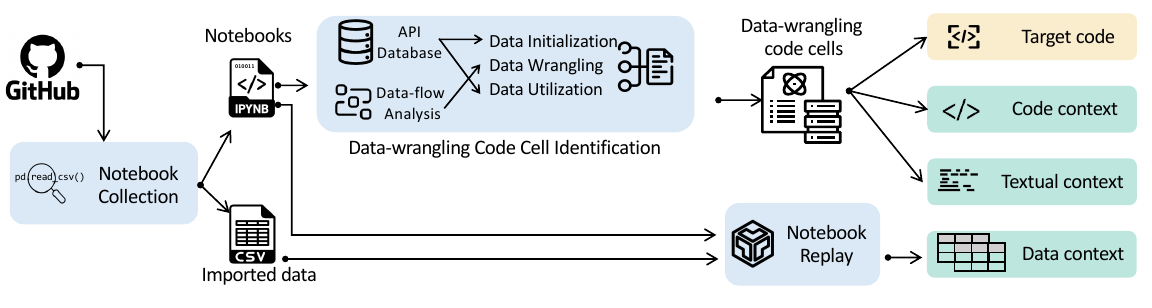}
\vspace{-2mm}
    \caption{The workflow of \mine.}
\label{fig:dataset:cocomine}
\end{figure*}

\section{\dataset Curation with \mine}
Mining data-wrangling code generation examples with sufficient context is non-trivial due to challenges of identifying data-wrangling scripts, collecting runtime data contexts, and distinguishing code and textual dependencies, as discussed in Section~\ref{sec:intro}. To address the challenges, we propose \mine to automatically collect data-wrangling code generation examples from public notebooks, resulting in a benchmark dataset called \dataset. 
The principle underlying \mine involves initially identifying data-wrangling code from tangled notebook cells, followed by the collection of contextual dependencies to construct examples.

Figure~\ref{fig:dataset:cocomine} shows the overall workflow of \mine. 
We first crawl notebooks relevant to data analysis and collect execution-needed data files used in notebooks from GitHub~(\S\ref{sec:dataset-collect-nb}). 
Then we parse the code in notebooks and conduct data-flow analysis to identify code cells related data-wrangling. These code cells track the lifecycle of a dataframe variable from initialization, wrangling, to final utilization in a data analysis task~(\S\ref{sec:dataset-dsp}). Specifically, we create API databases to signify the data initialization~(\S\ref{sec:dataset-dsp-initialization}) and utilization stages~(\S\ref{sec:dataset-dsp-utilization}), and then conduct data-flow analysis on the dataframe variable to track its intermediate wrangling process~(\S\ref{sec:dataset-dsp-wrangling}). 
With identified data-wrangling code cells, we proceed to build code generation examples with multi-modal context~(\S\ref{sec:dataset-cge}). Specifically, we first sample potential target code from code cells in the data wangling stage~(\S\ref{sec:dataset-cge-target-code}). Next, we trace back the data-flow diagrams to collect the dependent code and textual context from cells preceding the target code~(\S\ref{sec:dataset-cge-textual-context}). Subsequently, we create a sandbox environment to replay notebooks and record the dataframe during execution to obtain runtime data context~(\S\ref{sec:dataset-cge-data-context}). 
Finally, we perform human verification to ensure the quality of the dataset~(\S\ref{sec:dataset-human-verification}). 

\subsection{Collecting Notebooks and Data Files}\label{sec:dataset-collect-nb}
\mine collects notebooks from GitHub using public GitHub API~\cite{githubapi}. The notebooks are required to have the \texttt{.ipynb} file extension and a redistributable license, following~\cite{yan2020autosuggest}.We remove duplicated notebooks and those from forked repositories. We choose GitHub as our source, as opposed to Kaggle~\cite{kaggle}, a platform for data scientists to display their DS solutions through notebooks. This decision is based on the fact that GitHub offers a larger and more diverse collection of notebooks~\cite{mondal2023cell2doc}. Subsequently, we apply a straightforward heuristic to collect notebooks potentially relevant to DS. The notebooks are required to run with Python~3 and invoke a \texttt{pd.read\_csv()} function from the \texttt{Pandas} library, which is one of the most popular data analysis libraries in Python~\cite{pandas}. The utilization of \texttt{pd.read\_csv()} function indicates the loading of real data from external files, as opposed to synthetic data used only for experimental purposes~\cite{kery2018literate}, which suggests a higher likelihood of the notebooks being employed for real-world data analysis~\cite{biswas2022dsp}. 

Then \mine collects data files loaded in notebooks to support execution. Specifically, we parse the code merged from code cells into an Abstract Syntax Tree (AST). We then identify the string nodes and assess whether the string represents a valid file system path. After that we trace the paths in original GitHub repositories and download the corresponding data files. Notebooks that rely on inaccessible data sources are excluded. Finally, we simplify the nested file paths in the notebooks and consolidate all files into a single directory in preparation for execution.

\subsection{Data-Wrangling Code Cell Identification}\label{sec:dataset-dsp} 
In this step, \mine extract data-wrangling code cells from tangled notebooks. These code cells compose a self-contained data analysis process and track the lifecycle of a dataframe variable. 
We introduce three typical stages in this process: 
a data initialization stage to initialize a target dataframe, a data wrangling stage to convert the dataframe into a desired format, and a data utilization stage to utilize processed data for an analysis objective such as visualization and statistic computation. 
To identify the three stages of data-wrangling code cells, we create API databases and conduct data-flow analysis~\cite{wang2021restoring, biswas2022dsp, titov2022resplit}. The rationale of analysing API and data-flow stems from the fact that a majority of data-wrangling codes are implemented with common Python DS libraries such as \textit{pandas} and \textit{matplotlib}~\cite{biswas2022dsp}. These libraries offer high-level and stable APIs to perform specific tasks on data, which can be used to infer code functionality and subsequently deduce its stage.

Specifically, we parse the programs into an AST and extract the temporal sequence of API calls through standard static analysis. We then conduct API matching for each statement with pre-defined API databases to determine the initialization of a dataframe and its utilization. After that we conduct data-flow analysis of the target dataframe to identify its transformation stage. Code cells that undergo these states are extracted as the valid data-wrangling code cells. 
We keep the top-to-bottom order of code cells as in the original notebook to keep the sequential logic. 
To form a comprehensive API database, we empirically investigate API documentation, statistically count API frequencies, and also refer to previous studies~\cite{yan2020autosuggest, biswas2022dsp}, to collect the representative APIs. To ensure self-containment of DS tasks, we discard notebooks with no complex code cells extracted. The methods to identify each stage are described in the following section. 

\subsubsection{Data Initialization}\label{sec:dataset-dsp-initialization} 
A data analysis task starts with the \textit{data initialization} stage, whereby a dataframe variable is initialized. This variable serves as the target variable that we aim to scrutinize throughout the entire process. 
Specifically, we first identify statements for data initialization with API matching and then select assigned variable in the statement as the target variable.
We select four types of statements:
(1) Directly creating a new dataframe with the API, such as \texttt{df=pd.DataFrame()}.
(2) Loading a dataframe from external data files with the API, such as \texttt{df=pd.read\_csv()}.
(3) Manipulating an existing dataframe with the API to return a new dataframe, such as \texttt{df=pd.merge([df1, df2])}.
(4) Assigning a new dataframe via dataframe operations, such as slicing and addition.
Table~\ref{tab:API-target-var} shows the full API databases for data initialization.  It is noteworthy that APIs from other libraries can also return a dataframe (\eg \texttt{sns.load\_dataset(\textquotedbl iris\textquotedbl)} from \texttt{seaborn}). While our focus is on presenting a general identification framework, we have collected APIs from the widely-used \texttt{Pandas} library. Nevertheless, our approach can be readily extended by incorporating more APIs.

\begin{table}[t]
\centering
\caption{APIs and Operations to Identify Dataframe Variables.}
     \vspace{-2mm}
    \resizebox{0.44\textwidth}{!}{
    \begin{tabular}{cl}
    \hline
          Category & \multicolumn{1}{l}{API calls \& Operations} \\
    \hline
    \multirow{1}[1]{*}{Definition}  & pd.DataFrame(), pd.Series()   \\
    \hline
    \multirow{3}[2]{*}{\shortstack{Data\\Loading}}  
     & pd.read\_csv(), pd.read\_pickle(), pd.read\_excel()  \\
     & pd.read\_json(), pd.read\_table(), pd.read\_hdf()   \\
     & pd.read\_sql(), pd.read\_html() \\
    \hline
    \multirow{3}[2]{*}{\shortstack{Dataframe\\Manipulation}}  & pd.get\_dummies(), pd.merge(), pd.concat()   \\
     & pd.stack(), pd.unstack(), pd.cut(), pd.qcut()   \\
     & pd.melt(), pd.corr(), pd.crosstab(), pd.pivot\_table()  \\
    \hline
    \multirow{1}[1]{*}{Operations}  & +, -, *, /, df[$\cdot$]   \\
    \hline
    \end{tabular}%
    }
  \label{tab:API-target-var}%
\end{table}%

\subsubsection{Data Wrangling}\label{sec:dataset-dsp-wrangling} 
After initialization, the data need certain processing, which we categorize as the \emph{data wrangling} stage. This stage involves exploration, cleaning, and feature engineering, which aids in preparing the appropriate data for subsequent processing. Well-prepared data reduce the efforts for data analysis and contributes to the successful resolution of a data analysis task. 
Specifically, this stage can involve multiple statements to update the target variable. 
To identify wrangling codes, we conduct data-flow analysis to track the target variable defined in data initialization. A statement is designated as the data wrangling code if the target variable is reassigned to itself (\eg \texttt{df=df.dropna()}), or an in-place modification is invoked on itself (\eg \texttt{df.dropna(inplace=True)}).
Here, we track only one variable in the cells so that we can check whether data has been tampered with.
We do not consider the assignment to a new variable as the semantics can be changed when such an assignment is made.

\subsubsection{Data Utilization}\label{sec:dataset-dsp-utilization} 
The final data utilization stage is to complete a specific objective in DS, where analysts aim to reach a conclusion after rounds of programming and analysis. This stage is often achieved through the creation of visualizations or calculation of statistical measures with the well-transformed dataframes to facilitate future decision-making. 
Based on previous works~\cite{biswas2022dsp, kery2018literate} and our observations, these intentions can often be achieved through the use of well-developed API calls from popular DS libraries. Therefore, we can designate the statement as the DS objective through examining the API calls invoked on the target variable. 
We select four most frequently used libraries for visualization and statistics to create API databases, i.e., \textit{matplotlib}, \textit{seaborn}, \textit{scipy}, and \textit{scikit-learn}. 
Table~\ref{tab:API-final-goal} shows the full API databases for data utilization. Our findings of final objectives also align with the Biswas et al. that DS problems in Kaggle notebooks are mostly finished with a evaluation or visualization stage~\cite{biswas2022dsp}.

\begin{table}[htbp]
\centering
\caption{API Calls to Identify Data Utilization Statements. }
    \resizebox{0.4\textwidth}{!}{
    \begin{tabular}{cl}
    \hline
          Library & \multicolumn{1}{l}{API calls} \\
    \hline
    \multirow{3}[2]{*}{matplotlib}  
     & show(), plot(), barh(), hist(), hist2d(), imshow()  \\
     & bar(), pie(), scatter(), contour(), pcolormesh()   \\
     & errorbar(), matshow(), semilogy(), probplot()  \\
    \hline
    \multirow{7}[3]{*}{seaborn} 
     & distplot(), heatmap(), countplot(), barplot() \\
     & boxplot(), jointplot(), clustermap(), distplot()    \\
     & pairplot(), scatterplot(), violinplot(), kdeplot()     \\
     & swarmplot(), regplot(), rugplot(), stripplot()     \\
     & lmplot(), catplot(), relplot(), lvplot(), tsplot()     \\
     & lineplot(), boxenplot(), factorplot(), palplot()     \\
     & pointplot(), corrplot(), residplot(), FacetGrid()   \\
    \hline
    \multirow{1}[1]{*}{scipy}  & stats.plot(), stats.hist(), stats.scatter()   \\
    \hline
    \multirow{6}[2]{*}{scikit-learn}  
     & mean\_squared\_error(), mean\_absolute\_error() \\
     & roc\_auc\_score(), cross\_val\_score(), r2\_score()  \\
     & confusion\_matrix(), recall\_score(), f1\_score()   \\
     & pearsonr(), accuracy\_score(), precision\_score() \\
     & classification\_report(), cosine\_similarity() \\
     & softmax\_cross\_entropy\_with\_logits() \\
    
    \hline
    \end{tabular}%
    }
  \label{tab:API-final-goal}%
\end{table}%

\subsection{Multi-modal Context Alignment}\label{sec:dataset-cge}
Next, we create examples for contextualized code generation based on the extracted data-wranging code cells to construct \dataset dataset. As shown in Figure~\ref{fig:intro:task-definition}, given the multi-modal context of code, text, and input-output dataframe, the task is to generate the target code that can transform the input dataframe to the output dataframe. In the following, we first introduce the formulation of target code. Then we elaborate data context, code context, and textual context, and finally show the final example and dataset statistics. 

\subsubsection{Target Code}\label{sec:dataset-cge-target-code}
The target code is formulated as the entire code in code cells rather than several statements, as a code cell is a basic functional unit for programming and execution in notebooks. 
We use code cells between data initialization and data utilization stages to assure the functionality of data wrangling. 
We select target code cells based on the observation that DS programmers tend to perform data inspection operations during programming, such as displaying a dataframe with \texttt{df.head()} in the last line of a code cell. The inspection helps programmers to understand table details before programming or evaluate the code functionality once the programming is finished~\cite{kery2018literate, rule2018exploration}, which can be regarded as a signal for the beginning and functional completeness of a code snippets~\cite{chapman2024provenance}
Hence, we select the code between any two consecutive data inspection statements in the data-wrangling code cells as our target code. 
Specifically, we use the built-in inspection APIs to identify the data inspection statements, including the dataframe display expression (\ie~\texttt{df}), and partial dataframe display expressions (\ie~\texttt{df.head()} and \texttt{df.tail()}). 
It's worth noting that the target code can span multiple code cells; thus, we merge them to obtain the final target code. Figure~\ref{fig:data-stat-pie} shows the distribution of APIs and modules used in the target code.

\subsubsection{Data Context}\label{sec:dataset-cge-data-context}
Upon identifying the target code, we linearly re-execute the notebook to obtain runtime data context, \ie the input and output dataframes. Specifically, we record the value of the target variable before target code as the input dataframe, and then execute target code to obtain the output dataframe. In the installed sandbox, we sequentially execute code cells from top to bottom for simplicity without considering their original arrangement in notebooks. 
Notebooks that encounter exceptions or exceed execution duration of 300 seconds per cell are considered unsuitable, resulting in the subsequent exclusion of examples derived from them. 

\noindent \textbf{Execution Environment}
The challenge of executing real-world computational notebooks is the potential inclusion of diverse libraries and dependencies from other \texttt{.py} files. However, creating a customized environment for each notebook individually is impractical for large-scale execution due to the intricate dependencies and the resulting increase in evaluation time~\cite{wang2021restoring, wang2020assessing}.
To address this, we adopt a simplified approach by creating a unified sandbox environment for all notebooks in our dataset. 
This sandbox is equipped with Python standard libraries, specifically version 3.6.13. Additionally, we install the 15 most frequently used non-standard data analysis libraries in notebooks, namely: \textit{pandas}, \textit{numpy}, \textit{matplotlib}, \textit{sklearn}, \textit{seaborn}, \textit{scipy}, \textit{nltk}, \textit{plotly}, \textit{statsmodels}, \textit{geopandas}, \textit{bokeh}, \textit{ggplot},  \textit{xgboost}, \textit{lightgbm}, and \textit{patsy} based on our statistical analysis. 
Despite the popularity of machine learning libraries, such as \textit{TensorFlow} and \textit{PyTorch}, we do not install them due to their extensive usage in DL tasks, which require extensive time and computation resources, making them less suitable for execution evaluation

\subsubsection{Code Context}
In code generation, the code contexts prior to the target code are essential as they can provide clues for variable reusing and already performed computations~\cite{Iyer2018ConCode}. However, mining relevant code context is difficult, especially in the notebook environment, where cells at the very beginning of notebooks can be referred, such as import statements and imported dataframes~\cite{chen2021plotcoder}. To address the issue, we consider two types of code context. On one hand, we incorporate code cells in the data-wrangling code cells prior to the cell of target code, which specify the performed computations on the target variable. On the other hand, we incorporate the dependent statements of (1) the latest definition of variables used in the target code, and (2) the import statements of DS libraries, which can be used to infer variable and API usages. We identify the dependencies by matching module names and variable references with AST, which are then merged into a new code cell and placed at the beginning before other code cells.

\subsubsection{Textual Context}
In DS notebooks, natural language documentations are often well-written by data scientists to enable collaboration and sharing usage~\cite{wang2022documentation, mondal2023cell2doc, cui2022codeexp}, which are useful to provide task background and code instructions. 
Therefore, we include textual context to our task, which includes: 
(1) markdown cells before the target code cell and before each code cell in the data-wrangling code cells, 
(2) comments in the cells of code context, 
and (3) inline comments in the target code cells. We include the inline comments because they can specify (partial) functions of the target code, which aids in good-quality generation, as suggested by previous work~\cite{mondal2023cell2doc}.
One type of dependency is the natural language explanations in \textit{markdown cells}, which we can incorporate by inserting the original non-blank markdown cells before each code cell.

\begin{figure}[t]
\centering      
	\subfigure[Module counts.] %
	{
		\begin{minipage}{0.47\linewidth}
			\centering      
		    \includegraphics[width=1\linewidth]{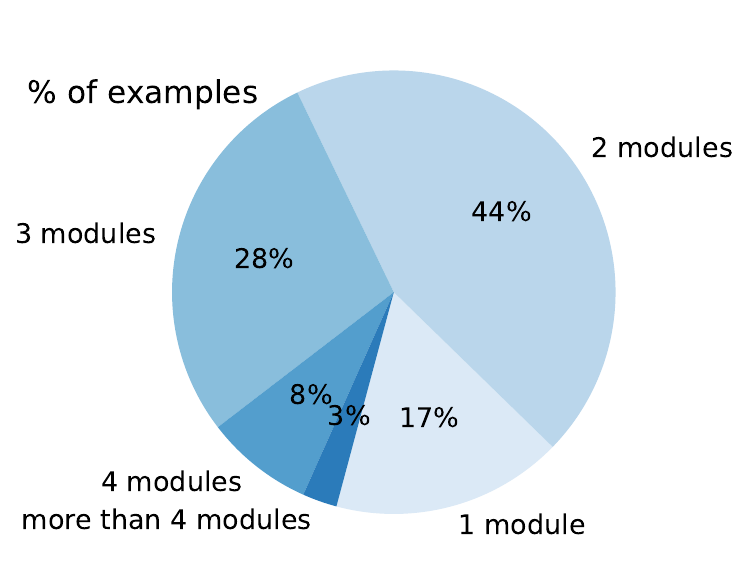}  
		\end{minipage}
	}
	\subfigure[API counts.] %
	{
		\begin{minipage}{0.42\linewidth}
			\centering    
			\includegraphics[width=1\linewidth]{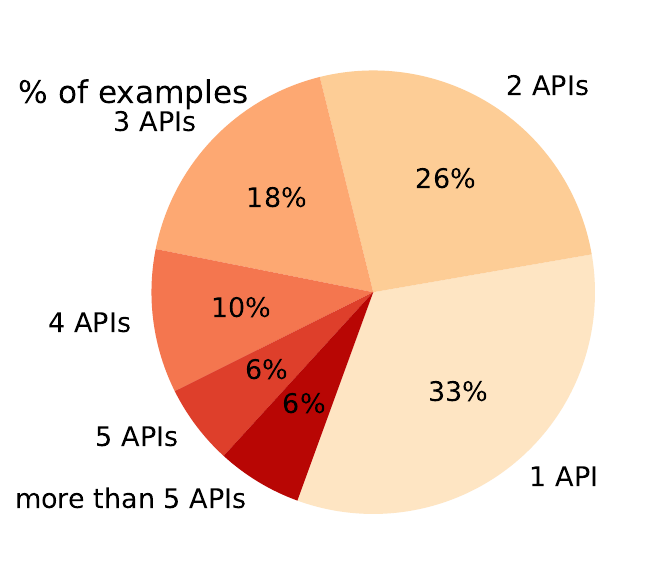}  
		\end{minipage}
	}
	\subfigure[Module frequency. ]  %
	{
		\begin{minipage}{0.45\linewidth}
			\centering      
		    \includegraphics[width=1\linewidth]{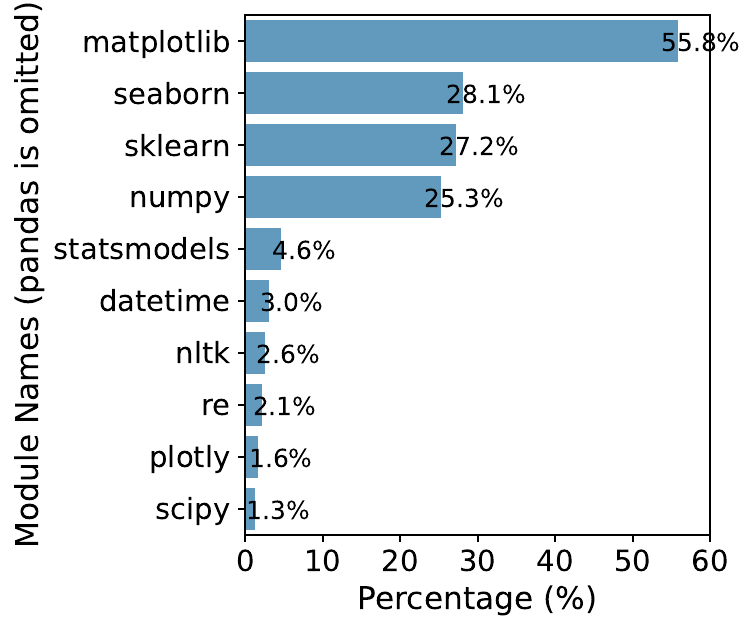}  
		\end{minipage}
	}
	\subfigure[API frequency. ] %
	{
		\begin{minipage}{0.47\linewidth}
			\centering     
			\includegraphics[width=1\linewidth]{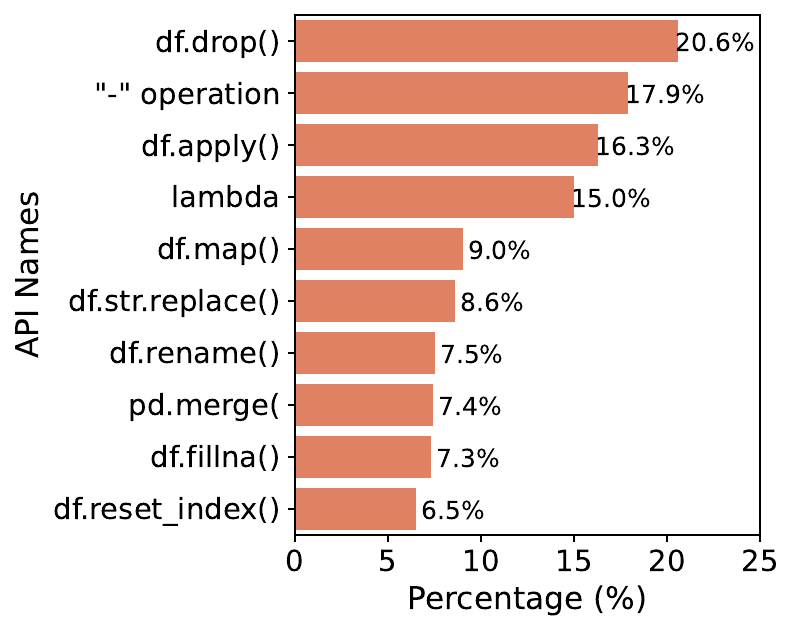}  
		\end{minipage}
	}
	
	\caption{Distribution of modules and APIs in the target code.} %
	\label{fig:data-stat-pie}  %
\end{figure}

\subsubsection{Finalizing Context}\label{sec:dataset-cge-filter}
Finally, to preserve sequential information, we merge code and textual contexts as a unified part and keep their original top-to-bottom order as in the notebook. 
To reduce task complexity, we discard examples with more than ten context code cells. We also remove those with identical input and output dataframes as this may indicate a vague functionality of target codes. 
For uniqueness, we remove duplicated examples and examples with the same target code. However, it's often the case that even in different notebooks, there are the same code snippets in the code cells, due to the convenience of modularity and reuse. To avoid answer leakage, we remove testset examples whose target code appears in the context of training and validation examples. 

The final \dataset contains 54,567 and 2,000 examples for training and validation and 1,654 for testing.
To ensure rigorous evaluation of correct output dataframes, we leverage the full data with numerous rows rather than the snapshots to producing output dataframes during testset execution. However, as using full dataframes in training is computationally expensive for modeling, we do not incorporate full dataframes and simply keep a maximum of ten rows in the training and validation set. 

\subsubsection{Dataset Statistics}\label{sec:dataset-cge-statistics}
Table~\ref{tab:data-constrcution-data-statitics} shows the statistics of \dataset. We find that test examples contain a lower average number of textual context tokens (138.0) and target code tokens (41.3) compared to the training and validation examples.
Figure~\ref{fig:data-stat-pie} shows the distribution of modules and APIs used in target code of \dataset. We find that most examples use less than 3 modules(89\%) and 3 APIs(77\%). Furthermore, we provide the frequency of API and module usage to elucidate task distribution, encompassing a broad spectrum of widely-used DS modules and data wrangling APIs.

\subsection{Human Verification}\label{sec:dataset-human-verification}
Here we conduct a human evaluation to verify the solvability of our task. We enlist the participation of three university students and two industry data scientists, all of whom possess extensive experience with Python and notebook environments. To ensure a rigorous and credible validation process, we formulate a comprehensive guidebook and conduct a pilot study on several examples following the guide. The guidebook includes our task objectives, evaluation criteria, a step-by-step annotation guide, and five representative examples. Next we randomly sample 50 examples from the testset as the final verification data for participants. 
Our annotation task is to judge if the provided information is sufficient for the participants to write target code, given varying combinations of contextual information, \ie, code, text, input dataframe (data$_{in}$), and output dataframe (data$_{out}$). During annotation, they are allowed to discuss and search the Internet to ensure annotation accuracy and quality. Finally, the class with the majority votes is assigned to each example. 
We find that 34\% examples can be solved with only the context of code$+$text, while 40\% and 42\% examples can be resolved with the code$+$text$+$data$_{in}$ or code$+$text$+$data$_{out}$, respectively. Furthermore, 76\% examples can be solved when all elements (code$+$text$+$data$_{in}$ $+$data$_{out}$) are accessible.
The results shows that the inclusion of both input and output dataframes significantly enhances the solvability rate. This is because participants can compare differences between input and output dataframes, thereby inferring the functionality of the target code.
These findings underscore the importance of integrating both input and output dataframes into the context of code generation. 
However, it is noteworthy that 24\% examples were challenging for the participants to evaluate, thereby highlighting the difficulty of our task.

\begin{table}[t]
  \centering
\caption{Statistics of \dataset Dataset.}
\resizebox{0.42\textwidth}{!}{
    \begin{tabular}{lrrr}
    \hline
    & \multicolumn{1}{l}{\textbf{train}} & \multicolumn{1}{l}{\textbf{dev.}} & \multicolumn{1}{l}{\textbf{test}} \\
    \hline
     \# examples & 54,567 & 2,000  & 1,654 \\
    \hline
     avg \# columns (input df) & 12.8  & 12.1  & 12.9 \\
     avg \# rows (input df) & 8.6   & 9.0   & 249.5 \\
     avg \# columns (output df) & 13.3  & 12.8  & 12.7 \\
     avg \# rows (output df) & 8.6   & 9.0   & 247.2 \\
    \hline
     avg \# textual context tokens & 276.2 & 283.9 & 138.0 \\
     avg \# target code tokens & 67.2  & 67.1  & 41.3 \\
    \hline
    \end{tabular}
}
\label{tab:data-constrcution-data-statitics}
\end{table}%

\section{Code Generation Methods}
In this section, we introduce models to perform contextualized data science code generation with \dataset. We comprehensively evaluate a diverse set of language models for code with varying sizes through finetuning (\S~\ref{sec:method-plm}) and prompting (\S~\ref{sec:method-llm}). To better incorporate data context, we also introduce \model with bi-encoders to encode runtime data context and textual context based on CodeT5 (\S~\ref{sec:method-biencoder}). The best model GPT-4 achieves 50.6\% accuracy, indicating a large room for improvement. 

\subsection{Finetuning Code Pretrained Models}\label{sec:method-plm}
Fine-tuning a pretrained language model (PLM) in downstream tasks is an effective paradigm for code intelligence~\cite{lu2021codexglue}. By training on example pairs, PLMs can transfer their knowledge to perform domain-specific code generation without learning from scratch~\cite{wang2022no}. 

\subsubsection{PLMs}\label{sec:method-plm-plm} We evaluate five representative code PLMs that have shown state-of-the-art performance in code generation. 

\textbf{CodeBERT} \cite{feng2020codebert} is an encoder-only architecture pre-trained on text-code pairs with masked language modeling and replaced token detection objectives. As the decoder is not pretrained, we apply a randomly initialized transformer decoder with 6 layers, 768 dimensions, and 12 heads, forming an encoder-decoder architecture to generate. The decoder is optimized during fine-tuning. 

\textbf{GraphCodeBERT} \cite{guo2020graphcodebert} enhances code representation by utilizing data flow graphs and uses an encoder-only architecture built on top of CodeBERT. Similar to CodeBERT, we train a decoder for generation using the same method.

\textbf{UniXcoder} \cite{guo2022unixcoder} is adapted from UniLM~\cite{unilm} and is pre-trained on unified cross-modal data including code, text, and AST. As UniXcoder is also an encoder-only architecture, we follow the method used in CodeBERT and train a decoder for generation. 

\textbf{PLBART} \cite{ahmad2021unified} is an encoder-decoder model built upon BART~\cite{lewis2020bart} and is pre-trained on a mix of Python functions, Java functions, and textual posts from StackOverflow~\cite{stackoverflow} with a denoising objective.

\textbf{CodeT5} \cite{wang2021codet5} is an encoder-decoder model built upon T5~\cite{raffel2020t5} and incorporates crucial token type information from identifiers and enables multi-task learning. The decoder is a multi-layer transformer with 12 layers, 768 dimensions, and 12 heads. 

\subsubsection{Finetuning Setting}\label{sec:method-plm-setting}
We prepare the model input by concatenating the code and textual context and input-output dataframes. 
We set the maximum sequence length of both textual context and data context to 256, resulting in a maximum concatenation length of 512. 
To ensure code syntax correctness, text from markdown cells is wrapped into multi-line comments by triple quotes. For data context, we concatenate the flattened string of input and output dataframes, where each table is converted by arranging each row's elements from left to right, then stacking them from top to bottom. Due to the input sequence length limit, we only use the first five rows of the input/output dataframes.
During fine-tuning, all models are trained on 8 Tesla V100 32GB GPUs with a cross-entropy loss function. We set the batch size as 64, the epoch as 20, and the maximum target sequence length as 300. The remaining hyper-parameters are kept by their default settings of each PLM.

\subsection{\model}\label{sec:method-biencoder}
The above PLMs encode the joint textual and data context, which may not optimal for this task due to two reasons. First, there is a discrepancy between textual and tabular data~\cite{huang2022otter}, which makes it challenging to understand the context and consider the cross-modality connections. Second, the input length limit of PLMs (\eg 512 tokens) often leads to incomplete input context, which can further hurt generation performance. 
To address the issues, we propose \model that leverages dual encoders to separately encode textual context and input-output dataframes, yielding more accurate latent representations for decoding. 

\begin{figure}[t]
    \centering
    \includegraphics[width=1\columnwidth]{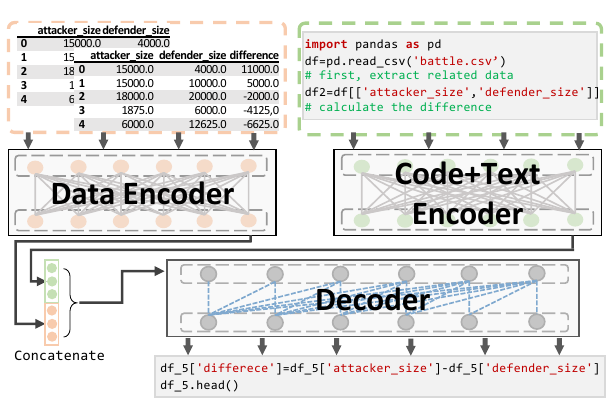}
    \caption{Model Architecture of \model.}
    \label{fig:mothod-datacoder}
\end{figure}

\subsubsection{Architecture}\label{sec:method-biencoder-arch}
The architecture of \model consists of two encoders and one decoder, as shown in Figure~\ref{fig:mothod-datacoder}. One encoder is used to encode code$+$text, and the other encodes tabular input/output dataframes. 
For each encoder, we apply $N$ transformer layers~\cite{vaswani2017attention} over the input sequences to produce text hidden states and table hidden states for code$+$text and dataframes, respectively. 
The input sequences are organized following the same setting described in~\S~\ref{sec:method-plm-setting}.
After encoding the input sequences, the decoder takes the concatenation of the last-layer hidden states from both encoders as input to generate target code. The decoder also applies $N$ transformer layers, but with a different attention mechanism, where tokens in each transformer layer can only attend to themselves and preceding tokens, allowing for sequential generation of the target code. 

\subsubsection{Training Setting}\label{sec:method-biencoder-setting}
To leverage pretrained knowledge and reduce training time, we initialize encoders and decoders with 12 layers, 768 dimensional hidden states, and 12 attention heads using the pre-trained parameters from CodeT5~\cite{wang2021codet5}. 
We use a learning rate of 2e-4, a batch size of 64, a maximum sequence length of 512 for both code$+$text and data, and a target sequence length of 300, respectively. We use AdamW optimizer~\cite{adamw} to fine-tune models for 20 epochs and perform early stopping on the development set.

\subsection{Prompting Large Language Models}\label{sec:method-llm}
Large language models (LLMs) have shown remarkable performance in various programming tasks such as code completion and generation~\cite{liu2024yourCodeCorrect}. Due to the high cost of tuning an LLM, we evaluate them through prompting, where a prompt with multi-modal context is constructed and then fed into LLMs to infer code~\cite{brown2020gpt2}.

\subsubsection{LLMs}\label{sec:method-llm-llm} We select four open-source code LLMs and three closed-source commercial LLMs to perform our task:

\textbf{StarCoder-15B} \cite{li2023starcoder} is a decoder-only model with 15b parameters built upon SantaCoder~\cite{allal2023santacoder}, employing Flash attention to scale up the context length to 8k. It is trained on The Stack dataset~\cite{kocetkov2022stack} including 86 programming languages, GitHub issues, Git commits, and Jupyter notebooks. 

\textbf{CodeLlaMa-34B} (CodeLlama-34b-Instruct) \cite{roziere2023codellama} is an autoregressive transformer model built upon Llama-2~\cite{touvron2023llama2}, using a multi-task objective of autoregressive and infilling prediction~\cite{bavarian2022efficient}, and long text fine-tuning to enable inference context length up to 16k. 

\textbf{Phind-CodeLlama-34B} (Phind-CodeLlama-34b-v2) \cite{phindcodellama} is an improved version of CodeLlaMa that are additionally trained on 1.5B tokens and 80k high-quality programming problems. 

\textbf{DeepSeek-Coder-33B} (deepseek-coder-33b-instruct) \cite{guo2024deepseek} is a transformer model trained from scratch on 2b tokens sourced from 87 programming languages. It employs a fill-in-the-blank task~\cite{bavarian2022efficient} with a 16K window to enhance code generation and infilling. 

\textbf{Codex} (code-cushman-001) \cite{chen2021evaluating} is an LLM with 12b parameters that powers GitHub’s Copilot auto-completion system. However, due to the advancements of GPT models for coding tasks, OpenAI will no longer support Codex API. 

\textbf{GPT-3.5} and \textbf{GPT-4}~\cite{chatgpt} are the recent state-of-the-art LLM developed by OpenAI. While they are not explicitly trained for code generation, they demonstrate notable performance in this domain~\cite{liu2024yourCodeCorrect}. Their effectiveness in handling code generation tasks is largely attributed to their large  parameter size. For GPT-3.5 and GPT-4, we use \textit{gpt-3.5-turbo-0613} and \textit{gpt-4-0613}, respectively. 

\begin{figure}[t]
    \centering
    \includegraphics[width=0.96\columnwidth]{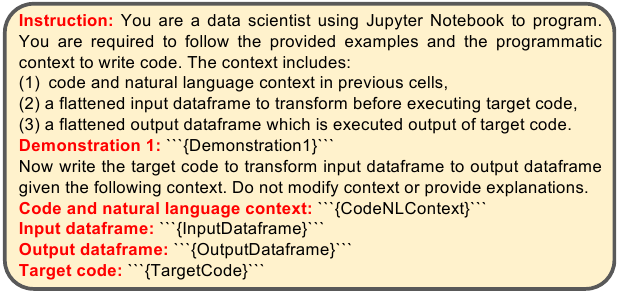}
    \caption{Prompt template to query LLMs.}
    \label{fig:method-prompt}
    \vspace{-2mm}
\end{figure}

\subsubsection{Prompting Setting}\label{sec:method-llm-setting}
We evaluate LLMs in both few-shot and zero-shot settings. 
Figure~\ref{fig:method-prompt} shows our few-shot prompt to query LLMs, which includes an instruction, demonstration, and context. 
To ensure code syntax correctness, the instruction, input dataframe and output dataframe are wrapped into multi-line comments by triple quotes.
In few-shot setup, we randomly sample $k$ in-context examples~\cite{brown2020gpt2, gao2023codeICL, li2024go} and insert them before the test example. 
This allows LLMs to learn from the in-context examples and generate corresponding response for the test sample. By default, we set $k$=2 for all LLMs. The zero-shot prompt is similar to the few-shot prompt, but without demonstrations.
We utilize the official OpenAI API~\cite{chatgpt} for Codex, GPT-3.5, and GPT-4, while the remaining open-source LLMs are based on the HuggingFace \texttt{transformers} library.

\subsection{Parameter-Efficient Finetuning}\label{sec:method-lora}
Besides small-sized code language models and large-sized ones, we also comprehensively evaluate medium-sized LLMs (1.3-7B) on \dataset through parameter-efficient finetuning (PEFT). 

\subsubsection{LLMs.} 
We select five medium-sized open-source code LLMs.

\textbf{DeepSeek-Coder-1.3B} and \textbf{DeepSeek-Coder-6.7B} (deepseek-coder-1.3b-base and deepseek-coder-6.7b-base) \cite{guo2024deepseek} are the DeepSeek-Coder family with varying parameter size, which are trained with the same datasets and training objectives.

\textbf{CodeGen-6B} (codegen-6B-mono) \cite{nijkamp2022codegen} is an transformer architecture and trained with the next-token prediction objective on three NL and code datasets sequentially: ThePile~\cite{gao2020pile}, BigQuery, and BigPython. 

\textbf{Mistral-7B-Instruct} (Mistral-7B-Instruct-v0.2) \cite{jiang2023mistral} is an  LLM tuned with instruction datasets based on Mistral-7B with grouped-query attention to accelerate inference speed.

\textbf{CodeLlama-7B-Python} \cite{roziere2023codellama} is a variant of CodeLlama LLM family built on top of Llama-2~\cite{touvron2023llama2}, using additional 100B Python tokens to  
specialize Python language.

\subsubsection{PEFT Setting.}
Following previous studies on PEFT for code generation~\cite{weyssow2023peftcode}, we adopt top-performing PEFT technique Low Rank Adaption (LoRA)~\cite{lora} to tune the LLM on \dataset. LoRA  freezes the model weights and injecting low-rank trainable matrices into the attention layers of the Transformer architecture, thereby drastically reducing the number of trainable parameters. We adopt the same input-output pairs as finetuning Code PLMs~(\S~\ref{sec:method-plm-setting}). For all LLMs, we use a decomposition rank of 8, a batch size of 3, and trained for 5 epochs on 1 Tesla A100 40GB GPU with a cross-entropy loss function. The remaining hyper-parameters are kept by their default settings of each LLM.

\section{Experiment}
We evaluate code generation models on \dataset by answering the following research questions (RQs):
\begin{itemize}[leftmargin=*, topsep=0pt]
    \item \textbf{RQ1:} How do different models perform on \dataset?
    \item \textbf{RQ2:} How do varying context affect performance?
    \item \textbf{RQ3:} How do various data context affect performance?
\end{itemize}

\begin{figure*}[t]
    \centering
    \includegraphics[width=0.96\linewidth]{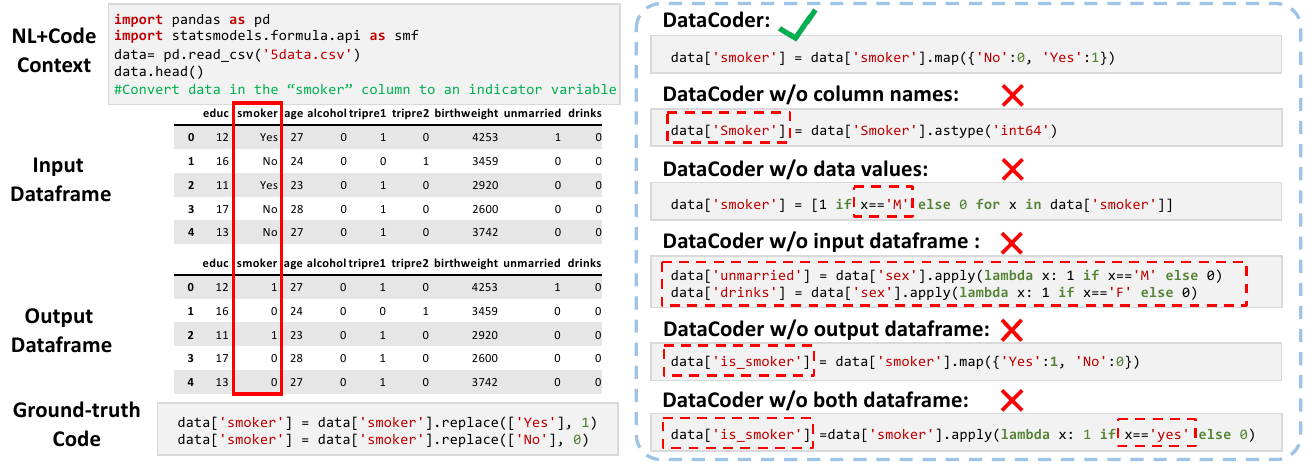}
    \caption{An example of predictions from models with diverse data context. Key errors are marked in red box.
}  
\label{fig:case1}
\end{figure*}

\subsection{RQ1: Performance of Different Models}\label{sec:experiment-main}
\noindent \textbf{Setup}: We first compare the performance of \model with a variety of PLMs and LLMs on \dataset. The generated code is evaluated using three metrics: \ie Exact Match (EM), CodeBLEU (CB), and Execution Accuracy (EA), as described in \S~\ref{sec:task-define-evaluation}. To ensure a rigorous evaluation, both PLMs and \model are finetuned on the identical training set of \dataset. For LLMs, in-context examples are sampled from the same training set. The results in terms of PLMs and LLMs are shown in Table~\ref{table:experiment-main-plm} and Table~\ref{table:experiment-main-llm}, respectively.

\begin{table}[h]
\small
\centering
\caption{Performance of \model and PLMs.
}
    \resizebox{0.44\textwidth}{!}{
      \begin{tabular}{l|ccc}
      \hline
      \hline
	\textbf{Model}&\textbf{EM}&\textbf{CodeBLEU}&\textbf{EA} \\
      \hline
       GraphCodeBERT& 14.0& 48.9 &28.9\\
       CodeBERT& 14.6& 49.0 &29.5\\
       UniXcoder& 15.1& 50.3 &30.7\\
      \hline
       PLBART& 16.3& 52.7 &33.1\\
       CodeT5& \underline{19.1} & 55.3 &38.1\\
      \hline
       \textbf{\model} (ours) & \textbf{21.3}& \underline{57.2} &\underline{42.2}\\
      \hline
      \hline
     \end{tabular}
    }
\label{table:experiment-main-plm}
\end{table}

\noindent \textbf{Results}: 
We first analyze the performance of \model in comparison with PLMs finetuned on \dataset. From Table~\ref{table:experiment-main-plm}, we find that:  
(1) Our \model, which leverages two separate encoders for textual and data context, outperforms all other PLMs across all metrics. This result shows the effectiveness of \model. 
(2) Compared with CodeT5, \model exhibits a consistent enhancement across all metrics  (\ie, +4.1\% EA, +2.2\% EM, and +1.8\% CodeBLEU). The improvement can be attributed to the use of a separate data encoder for input/output dataframes, which can incorporate a more extensive context and better represent the modalities to aid generation. 
(3) In the second group in \cref{table:experiment-main-plm}, due to the encoder-decoder architecture and denoising pre-training tasks being specifically designed for generation tasks, PLBART and CodeT5 perform better than encoder-only models such as CodeBERT, GraphCodeBERT, and UniXcoder.

\begin{table}[h]
\small
\centering
\caption{Performance of \model and prompting LLMs.
}
    \resizebox{0.44\textwidth}{!}{
      \begin{tabular}{l|ccc}
      \hline
      \hline
	\textbf{Model}&\textbf{EM}&\textbf{CodeBLEU}&\textbf{EA} \\
      \hline
	StarCoder-15B       & 8.1 & 53.2 & 33.2 \\
	CodeLlama-34B       & 1.0 & 36.9 & 4.9 \\
	Phind-CodeLlama-34B & 3.0 & 40.7 & 21.5   \\
	DeepSeek-Coder-33B  & 7.1 & 52.0 & 37.3  \\
      \hline
	  Codex           & 2.3& 31.5 &11.9\\
	GPT 3.5 & 6.9 & 55.3 & 40.1 \\
	GPT 4   & 11.1 & \textbf{60.8}  & \textbf{51.6} \\
      \hline
       \textbf{\model} (ours) & \textbf{21.3}& \underline{57.2} &\underline{42.2}\\
      \hline
      \hline
     \end{tabular}
    }
\label{table:experiment-main-llm}
\end{table}

Next, we analyze the performance of \model compared with prompting LLMs. From Table~\ref{table:experiment-main-llm}, we can find that: 
(1) Despite utilizing fewer parameters (305M), \model exhibits superior performance than five medium-sized code LLMs (12B\textasciitilde 34B) and one large-sized LLM (\ie GPT~3.5 with 175B). 
This can be attributed to our task requiring not only the comprehension of code and textual context but also the understanding of the data within input/output dataframes, which poses greater challenges in a few-shot setting. Fine-tuning on the training set proves beneficial in learning the ability to understand data context and consequently leads to improved performance. 
(2) Compared with GPT~4, although \model is inferior in terms of CodeBLEU and EA, it achieves a +10.2\% EM of the generated code. In addition, the finetuned PLMs consistently outperform in terms of EM. We attribute this phenomenon to the finetuning strategy applied to \model. As coding in notebooks for data analysis differs from general domain programming, current LLMs may not fully utilize domain knowledge and comprehend programming patterns without tailored finetuning, leading to more diverse code and lower EM score.

\begin{table}[h]
\small
\centering
\caption{Performance of \model and LLMs with LoRA.
}
    \resizebox{0.44\textwidth}{!}{
      \begin{tabular}{l|ccc}
      \hline
      \hline
	\textbf{Model}&\textbf{EM}&\textbf{CodeBLEU}&\textbf{EA} \\
      \hline
    Deepseek-Coder-1.3B & 17.2  & 59.5  & 45.2 \\
    CodeGen-6B & 17.8  & 60.2  & 43.9 \\
    Mistral-7B-Instruct & 16.8  & 59.8  & 46.1 \\
    CodeLlama-7b-Python & 19.0  & \underline{61.3}  & \underline{48.8} \\
    Deepseek-Coder-6.7B & \underline{19.3}  & \textbf{62.7}  & \textbf{50.5} \\
      \hline
       \textbf{\model} (ours) & \textbf{21.3}& 57.2 & 42.2 \\
      \hline
      \hline
     \end{tabular}
    }
\label{table:experiment-main-llm-lora}
\end{table}

Finally, we analyze the performance of LLMs that undergone tuning via LoRA. From Table~\ref{table:experiment-main-llm-lora}, we can find that: 
(1) Medium-sized models that have been fine-tuned consistently outperform \model in terms of CodeBLEU and EA metrics. We attribute it to increased parameter size and the expanded pretraining dataset size of these models, which significantly influence LLM efficacy as suggested by prevailing scaling laws~\cite{kaplan2020scaling}.
(2) Furthermore, comparing 15B-34B LLMs (Table~\ref{table:experiment-main-llm}) and tuning 1B-7B LLMs via LoRA, we find that tuning smaller models yields superior results compared to merely prompting larger models without tuning. Remarkably, the best DeepSeek-Coder-6.7B achieves comparable performance with GPT-4 (+8.2\% EM, +1.9\% CodeBLEU, and -1.1\% EA). This result implies the potential of parameter-efficient finetuning using a high-quality domain-specific dataset to enhance data-wrangling code generation in notebooks, which in turn underscores the utility of our dataset.

\subsection{RQ2: Effects of Varying Context }\label{sec:experiment-context}
\noindent \textbf{Setup}: To understand the importance of context in enhancing code generation, we conduct experiments by assembling various context to form the model input. We select two top-performing models to study, \ie the finetuned \model and GPT~3.5 with two in-context examples. We use the same split of dataset for evaluation.

\begin{table}[h]
\centering
\caption{Results with Different Programmatic Context.}
    \resizebox{0.48\textwidth}{!}{
    \begin{threeparttable}
    \begin{tabular}{l|ccc|ccc}
\hline
  	& \multicolumn{3}{c|}{\model}  & \multicolumn{3}{c}{GPT 3.5 (ICL $k$=2)} \\
	& \textbf{EM}  & \textbf{CB} & \textbf{EA}  & \textbf{EM}  & \textbf{CB} & \textbf{EA} \\
\hline
	  Full Context & \textbf{21.3}& \textbf{57.2} &\textbf{42.2} & \textbf{6.9} & \textbf{55.3} & \textbf{40.1} \\
        - Code+Text & 12.5& 47.8 &22.4 & 3.5 & 50.0 & 22.6 \\
        - Code+Text+InData & 14.1& 49.4 &23.8 & 3.3 & 50.4 & 25.0 \\
        - Code+Text+OutData & 18.4& 53.2 &34.5 & 4.6 & 52.0 & 29.4 \\
        - InData+OutData & 4.7 & 36.3 & 5.4 & 0.7 & 41.7 & 7.6 \\
        - InData+OutData+Text & 6.3  & 40.8 & 9.6 & 1.0  & 44.6  & 10.4 \\
        - InData+OutData+Code & 18.3 & 53.4 & 35.1 & 6.2 &  51.9 & 34.9 \\

\hline
	\end{tabular}
        \begin{tablenotes}
            \item[1] EM (Exact Match), CB (CodeBLEU), EA (Execution Accuracy).
            \item[2] InData (Input Dataframe), OutData (Output Dataframe).
        \end{tablenotes}
        \end{threeparttable}
	}
	\label{table:experiment-context}
\end{table}

\noindent \textbf{Results}: From Table~\ref{table:experiment-context}, we find that: 
(1) Removing data context (\ie \textbf{Code+Text}) leads to a substantial performance decline compared with \textbf{full context} (Code+Text+InData+OutData), indicating the indispensability of both dataframes for accurate inference of target codes.
(2) We also compare the importance of input and output dataframes by removing them separately (\ie \textbf{Code+Text+InData} vs. \textbf{Code+Text+OutData}). The results show that the output dataframe is more crucial in this task, as it typically includes all the information from the input dataframe in most cases (\eg cases involving the addition of a new column to the input dataframe). 
(3) Models struggle to generate correct code with only data context (\ie \textbf{InData+OutData}). This can be attributed to the inherent complexity of real-world programming-by-example tasks, where constraints such as variable names and APIs must be ensured for execution correctness. 
Moreover, the absence of runtime data contexts in the pretraining corpus of language models, as they are not typically saved in notebooks, further exacerbates the difficulty for current models to comprehend data context.
(4) Involving more natural language context (\ie \textbf{InData+OutData+Text}) marginally improves the performance over \textbf{InData+OutData}. This is possibly due to the frequent absence of textual instructions in notebooks since programmers do not necessarily write comments or descriptions when they program. 
(5) Incorporating code context (\ie \textbf{InData+OutData+Code}) proves more effective than textual context in aiding data wrangling code generation, suggesting the importance of code context in mitigating variable or API misuse.

\subsection{RQ3: Effects of Data Context}\label{sec:experiment-data-context}
\noindent \textbf{Setup}: In this RQ, we aim to understand the significance of data context in enhancing code generation. To answer this question, we first experiment by eliminating various components of the dataframe in the data context. Subsequently, we provide a varying number of table rows to further study the impacts of data values. Finally, we present a case study to intuitively show the impact of various inputs on the prediction results.

\begin{table}[h]
\centering
\caption{Results with Different Dataframe Components.}
    \resizebox{0.48\textwidth}{!}{
    \begin{threeparttable}
    \begin{tabular}{l|ccc|ccc}
\hline
  	& \multicolumn{3}{c|}{\model}  & \multicolumn{3}{c}{GPT 3.5 (ICL $k$=2)} \\
	& \textbf{EM}  & \textbf{CB} & \textbf{EA}  & \textbf{EM}  & \textbf{CB} & \textbf{EA} \\
\hline
	  full data context & \textbf{21.3}& \textbf{57.2} &\textbf{42.2} & \textbf{6.9} & \textbf{55.3} & \textbf{40.1} \\
        - w/o data values & 19.6& 55.3 &38.2 & 6.2 & 54.1 & 34.5 \\
        - w/o column names & 14.9& 52.9 &28.2 & 4.2 & 52.7 & 26.0 \\
\hline
	\end{tabular}
        \begin{tablenotes}
            \item[1] EM (Exact Match), CB (CodeBLEU), EA (Execution Accuracy).
        \end{tablenotes}
        \end{threeparttable}
	}
	\label{table:experiment-data-context}
\end{table}

\noindent \textbf{Results}: We investigate the effects of different components in the data context, \ie column names and data values. From Table~\ref{table:experiment-data-context}, we find that:
(1) Removing values within the input and output dataframes while retaining only the column names (\textbf{w/o data values}) results in a decrease of execution accuracy (EA) from 42.2\% to 38.2\% for \model and from 40.1\% to 34.5\% for GPT~3.5.
This result shows the importance of data values in enabling models to infer target code by comparing differences in values between input and output dataframes, particularly when code comments are absent.  
(2) Removing column names of dataframes (\textbf{w/o column names}) leads to a further performance decline, with EA dropping from 42.2\% to 28.2\% for \model and from 40.1\% to 26.0\% for GPT~3.5. This aligns with our intuition that the model needs to be aware of the existing columns in the data and which columns in the input dataframe contribute to producing the output dataframe; otherwise, the generated code might result in a key error and fail to execute.

\noindent \textbf{Case Study}:
In the example depicted in Figure~\ref{fig:case1}, the comment \textit{``Convert the data in the smoker column to an indicator variable''} provides crucial semantic information about the intended function of the target code. However, the model needs to discern which values in the ``smoker'' column should be mapped to which indicator variables and how to name the transformed columns. This information can be obtained by comparing the differences between the input and output dataframes. 
After using all information from both dataframes, we can see that \model generates source codes that correctly process the input dataframe to produce the output dataframe.
However, the models that are provided with partial data context fail to generate correct code. For example, without column names, the model mis-spells the key \textit{``smoke''} to \textit{``Smoke''}. Also, without data values, the model is unaware of the original flag (\ie \textit{``Yes''} and \textit{``No''}), and a flag \textit{``M''} is predicted. 
Eliminating input or output dataframes also leads to incorrect results. For example, without input dataframes, the model hallucinates the value to be converted. Without output dataframe, the model mis-spells the expected key \textit{``smoke''} to \textit{``is\_smoker''}. 
This example demonstrates the importance and data context and the effectiveness of our model.

\section{Related Work}

\subsection{Code Generation with Complex Context}
Code generation is a longstanding challenge in software engineering and natural language processing~\cite{Le2020ACMSurvey, roziere2023codellama, huang2024your}. 
Recent advances in large-scale pre-training have enabled code generation models to handle diverse tasks, which can generate target code with complex contexts in diverse modalities~\cite{nijkamp2022codegen}.  %
Textual context in natural language, such as human query~\cite{Hendrycks2021APPS, Austin2021MBPP, Zhu2022XLCoST, huang2021cosqa}
or comments~\cite{Yin2018CoNaLa, Lin2018NL2Bash, Wang2022MCoNaLa, Shi2022NaturalLT, li2023exploring}, can specify expected functions, which serves as high-level instructions in code generation~\cite{Oda2015LearningTG}. 
In complement, code context~\cite{kanade2020py150}, such as the code a programmer already written~\cite{Iyer2018ConCode, lu2021codexglue,li2024enhancing}, are also useful as they can provide detailed instructions, such as code logic and variable names. 
There are also works focusing on multi-modal code generation combining text and code~\cite{chen2021evaluating, zhou2022doccoder, wang2022odex, wang2022recode}. 
Apart from text and code, data context is another useful modality. 
Text-to-SQL generation~\cite{Zhong2017WikiSQL, Yu2018Spider} aims to generate SQL queries based on textual intents and sheets retrieved from databases. In this task, data context enables model to utilize important information, such as sheet names and column names, which can facilitate utilizing the correct data~\cite{katsogiannis2023text2sqlSurvey}.
Programming-by-example is another challenging task \cite{Pattis1994Karel, parisotto2017flashfill, le2014flashextract, barowy2015flashrelate, Polosukhin2018NeuralPS, chen2021latent}, where the specification is input-output examples. There are also efforts taking a step further that combine textual intents and input-output dataframes to generate table transformation code~\cite{jain2022jigsaw, khatry2023words2code}. 
Our work aims to support multi-modal code generation with an emphasis on leveraging all above three modalities, \ie code, text, and data, for better code generation.

\subsection{Data Science Code Generation } 
Code generation is important in automating low-level DS tasks\cite{Wang2021AutoDSTH}. 
Numerous models have been developed to synthesize code for various stages of the DS lifecycle, such as data preparation~\cite{bavishi2019autopandas, yan2020autosuggest, chopra2023cowrangler}, modeling~\cite{shi2022tfcoder, nam2022predictive}, and visualization~\cite{chen2021plotcoder, bavishi2021vizsmith, wu2020b2}.
As data wrangling tasks consume a significant amount of time for data scientists~\cite{Gulwani2016PBEData}, our focus is on automating data wrangling through code generation.

Computational notebooks is a popular programming tool in DS~\cite{subotic2022static, li2023edassistant}. JuiCe~\cite{agashe2019juice} is the first dataset to facilitate interactive code generation within notebooks, which inspired the development of various models \cite{chen2021plotcoder, Dong2022CODEP}. But these models are evaluated using surface-form metrics that cannot fully assess code quality as perceived by human developers. As a result, recent work has turned to execution-based metrics \cite{Chandel2022JuPyT5, Huang2022ExeDS, Yin2022DSN, Lai2022DS1000} for code generation evaluation. However, these studies omit the importance of data context and only provide a testset for evaluation, which cannot be applied to contextualized code generation.

\section{Conclusions and Future Work}
In this work, we introduce a novel task of contextualized data-wrangling code generation in notebooks with multi-modal context of code, text  and tabular input-output data. To mine examples for this task, we propose \mine that first extracts code cells relevant to data wrangling and then collects code generation examples with aligned multi-modal context, resulting in a dataset named \dataset with 58,221 examples to support training and execution-based evaluation. To better leverage input-output data, we propose \model with separate encoders for code and text contexts and input-output dataframes to generate target code.
Our experiments on a range of PLMs and LLMs verify the importance of data context and the effectiveness of our model.

In the future, we will explore the following directions: (1) notebooks involves runtime artifacts in a range of forms, \eg lists, images, machine learning models, and we will try to investigate the effects of these context in DS code generation; (2) automatic documentation in notebooks is also an important topic and we will investigate contextualized code documentation in notebooks; (3) there remains a large room for improvement and we will explore more powerful methods to perform this task.

\section{Acknowledgement}\label{sec:Acknowledgement}

We thank all reviewers for their valuable comments and suggestions. The work described in this paper was supported by the Research Grants Council of the Hong Kong Special Administrative Region, China (No. CUHK 14206921 of the General Research Fund).

\balance
\bibliographystyle{ACM-Reference-Format}
\bibliography{reference}

\end{document}